\documentclass{article}
\usepackage{natbib}
\usepackage{graphics}
\usepackage{graphicx}
\usepackage{latexsym}
\usepackage{a4}
\usepackage{a4wide}
\begin{document}

\hspace*{10.0cm}\parbox{4.0cm}{
KEK Preprint 2000-118 \\
NWU-HEP 2000-01 \\
DPNU-00-35 \\
TUAT-HEP 2000-5 \\
OCU-HEP 2000-01 \\
}

\begin{center}
\begin{Large}
\begin{bf}
Measurement of the cross-section and forward-backward charge asymmetry for 
the b and c-quark in $e^{+}e^{-}$ annihilation with inclusive muons 
at $\sqrt{s} = 58$ GeV
\end{bf}
\end{Large}
\end{center}

\begin{center}
TOPAZ Collaboration
\end{center}



\begin{center}
Y. Inoue$^{1}$, 
A. Miyamoto$^{2}$,
E. Nakano$^{1}$,
T. Takahashi$^{1}$,
T. Tauchi$^{2}$,
Y. Teramoto$^{1}$,
K. Abe$^{3}$,
T. Abe$^{3}$\footnote{Present address: Stanford Linear Accelerator Center, 
Stanford University, Stanford, California 94309, U.S.A.},
I. Adachi$^{2}$,
K. Adachi$^{4}$,
M. Aoki$^{3}$,
M. Aoki$^{5}$,
R. Enomoto$^{2}$\footnote{Present address: Institute for Cosmic Ray Research, 
University of Tokyo, Chiba 277-8582, Japan},
K. Emi$^{2}$,
H. Fujii$^{2}$,
K. Fujii$^{2}$,
T. Fujii$^{6}$$^{9}$,
J. Fujimoto$^{2}$,
N. Fujiwara$^{4}$,
H. Hirano$^{5}$,
B. Howell$^{7}$,
H. Hayashii$^{4}$,
N. Iida$^{2}$,
H. Ikeda$^{2}$,
S. Itami$^{3}$,
R. Itoh$^{2}$,
H. Iwasaki$^{2}$,
M. Iwasaki$^{4}$\footnote{Present address: University of Oregon, 
Eugene, OR 97403, U.S.A.},
R. Kajikawa$^{3}$,
K. Kaneyuki$^{5}$,
S. Kato$^{2}$,
S. Kawabata$^{2}$,
H. Kichimi$^{2}$,
M. Kobayashi$^{2}$,
D. Koltick$^{7}$,
I. Levine$^{7}$,
H. Mamada$^{8}$,
K. Miyabayashi$^{4}$,
K. Muramatsu$^{4}$,
K. Nagai$^{9}$\footnote{Present address: Queen Mary and Westfield College, 
Univrersity of London, London, E1 4NS,UK},
K. Nakabayashi$^{3}$,
M. Nakamura$^{1}$,
S. Noguchi$^{4}$,
O. Nitoh$^{8}$,
A. Ochi$^{5}$,
F. Ochiai$^{10}$,
N. Ohishi$^{3}$,
Y. Ohnishi$^{2}$,
Y. Ohshima$^{5}$,
H. Okuno$^{2}$,
T. Okusawa$^{11}$,
E. Shibata$^{7}$,
A. Sugiyama$^{3}$,
H. Sugiyama$^{4}$,
S. Suzuki$^{3}$,
K. Takahashi$^{2}$,
T. Tanimori$^{5}$\footnote{Present address: Department of Physics, 
Kyoto University, Kyoto 606-8502, Japan},
M. Tomoto$^{3}$,
T. Tsukamoto$^{2}$,
T. Tsumura$^{8}$,
S. Uno$^{2}$,
Y. Watanabe$^{5}$,
A. Yamamoto$^{2}$, and 
M. Yamauchi$^{2}$
\end{center}

\begin{center}
\begin{it}
$^{1}$Institute for Cosmic Ray Physics, Osaka City University, 
Osaka 558-8585, Japan 

$^{2}$KEK, High Energy Accelerator Research Organization, Tsukuba, 
Ibaraki 305-0801, Japan

$^{3}$Department of Physics, Nagoya University, Nagoya 464-8601, Japan

$^{4}$Department of Physics, Nara Women's University, Nara 630-8506, Japan

$^{5}$Department of Physics, Tokyo Institute of Technology, 
Tokyo 152-8551, Japan

$^{6}$Department of Physics, University of Tokyo, Tokyo 113-0033, Japan

$^{7}$Department of Physics, Purdue University, West Lafayette, IN 47907, USA

$^{8}$Department of Applied Physics, Tokyo University of 
Agriculture and Technology, Tokyo 184-8588, Japan

$^{9}$Department of Physics, The Graduate School of Science 
and Technology, Kobe University, Kobe 657-8501, Japan

$^{10}$Department of Physics, Faculty of Liberal Arts, 
Tezukayama Gakuin University, Nara 631, Japan

$^{11}$Department of Physics, Osaka City University, Osaka 558-8585, Japan

\end{it}
\end{center}

\begin{abstract}
  We have studied inclusive muon events  
using all the data collected by the TOPAZ detector at $\sqrt{s}=58$GeV
with an integrated luminosity of 273pb$^{-1}$. From 1328 
inclusive muon events, we measured 
the ratio $R_{q\bar{q}}$ of the cross section for $q\bar{q}$ production to the 
total hadronic cross section and forward-backward asymmetry $A_{FB}^{q}$ for 
b and c quarks. The obtained results are 
$R_{b\bar{b}}=0.13\pm0.02(stat)\pm0.01(syst)$,
$R_{c\bar{c}}=0.36\pm0.05(stat)\pm0.05(syst)$,
$A_{FB}^{b}=-0.20\pm0.16(stat)\pm0.01(syst)$ 
and $A_{FB}^{c}=-0.17\pm0.14(stat)\pm0.02(syst)$,
in fair agreement with a prediction of the standard model.
\end{abstract}



\section{Introduction}
\label{sec:introduction}

The $e^{+}e^{-} \rightarrow q\bar{q}$ cross section and
charge asymmetry for heavy quarks (b and c quarks) are fundamental 
quantities of electroweak interactions.
Especially, in the TRISTAN energy region, the maximum forward-backward charge 
asymmetry for quark pair production, predicted by the standard model, provides 
high sensitivity to quark couplings. 
In this paper, we report the final result of an analysis for the 
quark pair production cross-section ratio to the total hadronic 
cross-section ($R_{q \overline{q}}$) and 
forward-backward asymmetry of b and c-quark ($A_{\rm FB}$), 
using the data of all the high-luminosity runs collected by the TOPAZ detector 
at the TRISTAN electron positron collider. 
The data were collected from 1990 to 1995 
at $\sqrt{s} = 58$ GeV, corresponding to a total integrated 
luminosity of 273 pb$^{-1}$.
Previously, by using parts of the TOPAZ data, we reported on measurements 
of the heavy quark production cross section and charge asymmetry: 
through inclusive muons\cite{shimonaka}, 
inclusive electrons\cite{nakano,nagai} 
and $D^{*\pm}$\cite{nakano2}. 
In the previous inclusive muon analysis, we derived 
the b-quark parameters assuming the standard-model parameters for the c-quark
with 41 pb$^{-1}$ of data. 
The obtained result was $A^{b}_{\rm FB}=$ $-0.71\pm$0.34(stat)
$^{+0.07}_{-0.08}$(syst)\cite{shimonaka}.
Through inclusive electrons, the results were $A^{c}_{\rm FB}=$ $-0.49\pm$0.20
(stat)$\pm$0.08(syst) and 
$A^{b}_{\rm FB}=$ $-0.64\pm$0.35(stat)$\pm$0.13(syst), 
using 197 pb$^{-1}$ of data\cite{nakano}. 
In this paper, we derive $R_{q \overline{q}}$
and $A_{\rm FB}$ for both the b- and c-quark using the
inclusive muon events with improved statistics. Data samples in previous 
analyses are included. 
In addition, we perform a correction on the pion punch-through rate
using the measured data of pions from $\tau$ pairs.

The paper is structured as follows. A description of the TOPAZ detector and 
an overview of the data taking are given in Section \ref{sec:detector}. 
The hadronic event 
selection and the muon identification are described in Section 
\ref{sec:inclusivemuon}. 
Section \ref{sec:analysis} describes the analysis, 
including a Monte-Carlo simulation, 
flavor separation, fitting method, and 
systematic uncertainty of the measured parameters. 
Section \ref{sec:discussion} presents 
discussions of the fitting results and shows some figures compared to 
the other experimental results. Finally, conclusions are given in 
Section \ref{sec:conclusion}.

\section{Detector and data taking}
\label{sec:detector}
\subsection{TOPAZ detector}
\label{sec:detector-topaz}

\begin{figure}
\input epsf
\begin{center}
\leavevmode
\epsfxsize=10.0cm
\epsfbox{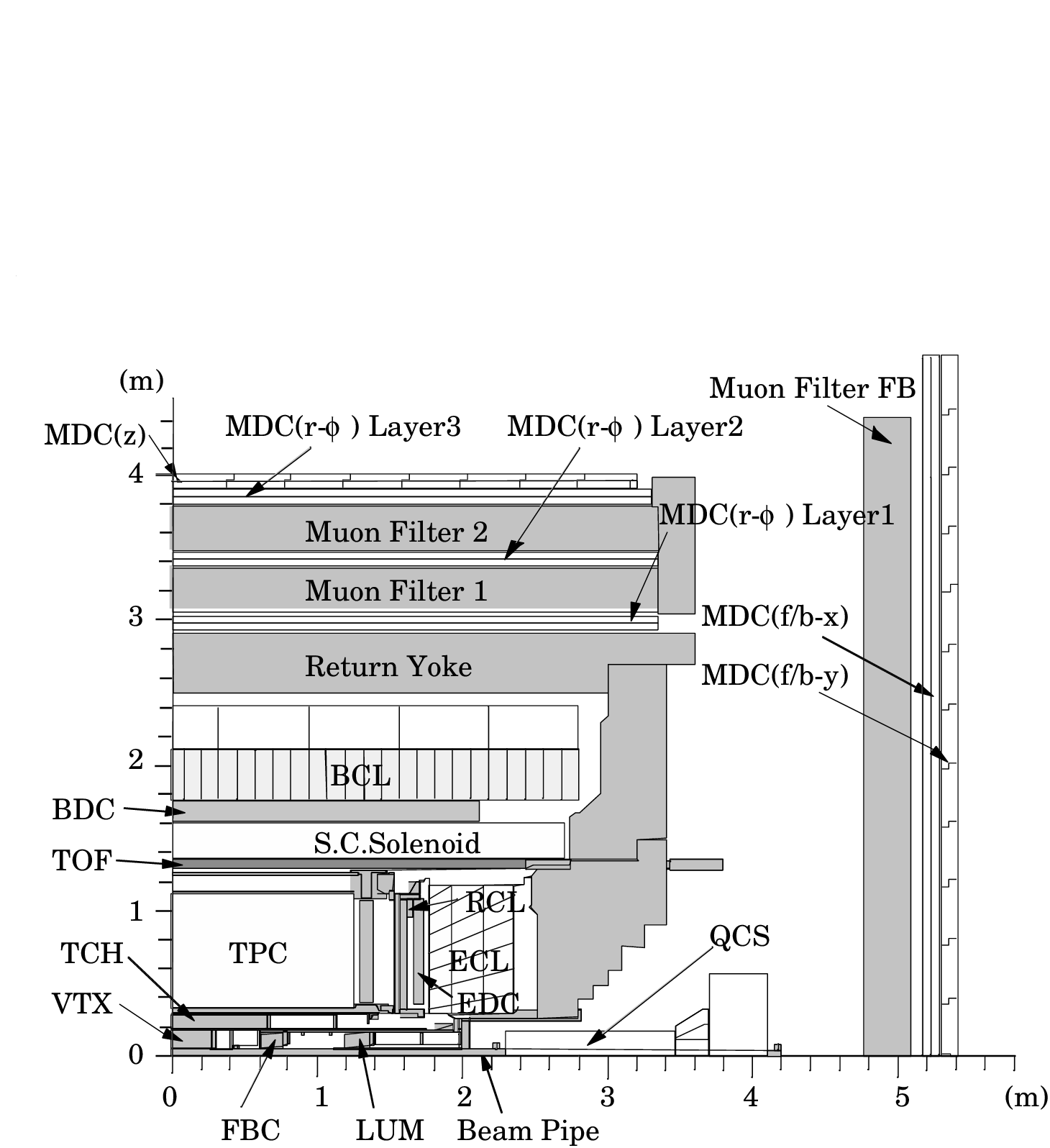}
\caption{Quadrant cross section of the TOPAZ detector. }
\label{fig:TOPAZ}
\end{center}
\end{figure}

TOPAZ detector was a general-purpose 4$\pi$ detector located
at the TSUKUBA experimental hall. 
A quadrant cross section of the TOPAZ detector is shown in 
Figure \ref{fig:TOPAZ}. The detector was 
upgraded in time for the high-luminosity runs which started in 1990. 
The upgrade was done by adding a vertex chamber,
a ring calorimeter and forward-backward muon chambers; further, the
inner drift chamber was replaced by a trigger chamber, and the luminosity
monitor was replaced by a forward calorimeter. After the upgrade, 
tracking of charged particle was done by a vertex chamber (VTX)
\cite{VTX}, a trigger chamber (TCH) and a time projection 
chamber (TPC)\cite{TPC}, which were placed inside of a 1T 
magnetic field, produced by a super-conducting 
solenoid magnet (SCS)\cite{SCS}. Time-of-flight counters (TOF)
\cite{TOF} placed inside of SCS were used to provide information about the 
time-of-flight and trigger. The energies of electrons 
and photons were detected with a barrel calorimeter (BCL)\cite{BCL}, 
a ring calorimeter (RCL), an endcap calorimeter (ECL)\cite{ECL,ECL2} 
and a forward calorimeter (FCL)\cite{FCL}. 
Those calorimeters were installed outside of the tracking devices. 
The total angular coverage of these calorimeters was $|\cos\theta| < 0.998$. 
The muon detection system (MDC)\cite{MDC} consisted of the barrel part and 
the forward-backward part. They were placed at the outermost part of the 
TOPAZ detector. The coordinate system used was: $z$ for the direction 
of the electron beam,
$\theta$ for the polar angle measured from the $z$ axis, $\phi$ for 
the azimuth angle measured from the horizontal direction 
pointing to the outward direction of the accelerator ring and $r$ for the 
radial direction from the beam axis. In this analysis, we mainly used TPC, BCL,
ECL and MDC.

\paragraph{Time projection chamber (TPC)}

The Time Projection Chamber (TPC) was the central tracking device.
Its function was to measure the 3-dimensional positions and dE/dX of charged 
particles. The inner and outer radii of the TPC were 36 cm and 109 cm, 
respectively, and its length was 260 cm. Its polar angle coverage was 
$| \cos\theta |<0.83$. 
They were made of 8 sectors divided into the forward and backward side, 
separated by a central membrane. 
The detection of drifted electrons was performed by 16 multiwire 
proportional chambers, placed on the end plane of each sector. 
The chamber was filled with an Ar/CH{\tiny 4} = 90/10 mixture 
gas (P10) at 3.5 atm. The momentum resolution of TPC was 
$\sigma_{p_{T}}/p_{T} = 
\sqrt{(1.5p_{T}({\rm GeV/c}))^{2} + (1.6)^{2}}$\%. 

\paragraph{Electromagnetic calorimeter (BCL and ECL)}

The barrel calorimeter (BCL) was made of 4,300 lead glass blocks with 
photo-multipliers. Each lead glass block had a 20 radiation length. 
Its polar-angle coverage was $|\cos\theta| < 0.84$. The BCL energy resolution 
was $\sigma_{E}/E = \sqrt{(8.0)^{2}/E({\rm GeV}) + (2.5)^{2}}$\%,
measured by Bhabha events. 

The endcap calorimeter (ECL) was a pair of sampling gas calorimeters, and 
it consisted of a sandwich of conductive plastic tubes with 
cathode readouts and 2 or 3 mm thick lead plates. The total radiation length 
of the detector was 18. Its coverage 
was $0.85<|\cos\theta|<0.98$. The energy resolution of ECL was 
$\sigma_{E}/E = 6.7$\% for Bhabha events.

\paragraph{Muon drift chamber (MDC)}

The muon detector (MDC) consisted of a barrel part (BMU) and a 
forward-backward part (EMU). In the present analysis, we only used BMU.
BMU was located outside of the return yoke, which was made
of 40 cm thick steel. BMU consisted of 3 superlayers of muon chambers 
interleaved with 2 layers of 30 cm thick muon filters made of steel. Each 
superlayer had double-layers of drift chambers, both measuring the $\phi$ 
direction 
of the track. 
In the outermost superlayer, we had an extra double-layer of drift chambers 
that measured the z direction of the track.
The polar-angle coverage of BMU was 
$|\cos\theta|<0.66$, and the minimum absorption length of BMU was 7.9. 
The muon drift chambers were made of extruded aluminum tube modules.  
One module contained eight cells, each 10 cm wide and 5 cm high, 
arranged in two layers of four cells each, staggered by a half cell 
to cover the dead regions near the cell walls. 
The averaged detection efficiency of each plane was 96\%, measured 
using cosmic rays. 
 
\paragraph{Trigger system} 
The main triggers were made of an energy trigger, a track trigger 
and a muon trigger\cite{Trigger,Soft-trig,Muon-trig}. 
The majority of the hadronic events were triggered by the energy trigger. 
The energy trigger required one of 
the following four conditions to be satisfied: 1) The total energy 
deposited in BCL be greater than 2 or 4 GeV, 
depending on the run conditions, or 2) the sum of the energy in the 
forward and backward ECL detectors be greater than 10 GeV, 
or 3) the BCL system had two energy clusters, 
both greater than 1 GeV, or 4) there were no energy clusters of energy 
greater than 3 GeV in neither the forward nor the backward ECL.   
The rate of energy triggering was approximately 1 Hz. 

The track trigger required that there be at least 2 tracks having an 
opening angle larger than 45$^{\circ}$ in the $r-\phi$
plane. Their vertex position was required to be within $\pm$ 20 cm of the 
interaction point along the beam direction by a software trigger. 

The muon trigger system was implemented for triggering 
$e^{+}e^{-}\rightarrow\mu^{+}\mu^{-}$ events in high-luminosity runs.
Muon track signals were made of 8-OR signals of BMU drift tubes in coincidence 
with TOF signals. The muon trigger required two muon track signals. The opening
angle of two tracks was required to be greater than 135$^{\circ}$. 

\subsection{Data taking}

TOPAZ started data taking in the spring of 1987 and ended 
in the summer of 1995. The total integrated luminosity was 340pb$^{-1}$.
In 1990, the TRISTAN main ring was upgraded so as to increase 
the luminosity. Then, the collision energy was fixed at 58 GeV 
to maximize the luminosity. 
The data used in this analysis were taken at $\sqrt{s} = 58$ GeV during 
the period given in Table \ref{tab:DATA}. 
The data before 1990 and the data in January 1993 were not used
because the beam energy was not 29 GeV. The data in the first quarter of 
1995 were not used because of a detector problem.

\begin{table}[h]
  \caption{Event sample.}
  \label{tab:DATA}
  \begin{center}
    \begin{tabular}{lrr}
      \hline \hline
	Period& integrated luminosity (pb$^{-1}$)& No. of Hadronic events\\
	\hline
	Feb, 1990$\sim$Dec, 1992 & 114& 12,811\\
	Feb, 1993$\sim$Dec, 1994 & 144& 14,942\\
	Apr, 1995$\sim$May, 1995 & 15 & 1,808\\
      \hline
                               & 273& 29,561\\
      \hline \hline
    \end{tabular}
  \end{center}
\end{table}

\section{Inclusive muon selection}
\label{sec:inclusivemuon}

\subsection{Hadronic event selection}
\label{sec:inclusivemuon-selection}

 To select inclusive muon events, we first selected hadronic
events from DSTs using the TOPAZ standard selection criteria for 
hadronic events \cite{hadsel}, which were:

\begin{itemize}

\item[(1)]At least five ``good'' tracks were coming from the interaction point,
where a ``good'' track was defined by (i) $r<5$ cm and $|z|<5$ cm at the closest 
point of approach to the beam axis, (ii) $p_{T}>0.15$ GeV/c, and (iii) 
$|\cos\theta|<0.83$. 

\item[(2)]The total visible energy, $E_{vis}$, had to exceed the beam energy,
$E_{beam}$.

\item[(3)]The momentum balance,$|\sum p_{z}|/E_{vis}$, was less than 0.4.

\item[(4)]The larger of the invariant jet masses in the two hemispheres, 
divided by the plane perpendicular to the thrust axis, $M_{\rm jet}$, 
exceeded 2.5 GeV/$c^{2}$.

\item[(5)]The number of clusters having an energy greater than $E_{beam}/2$ was 
less than 2. 

\end{itemize}

The background processes for the hadronic events were mainly 
$\tau$ pairs and two-photon events. Especially, $\tau$ pairs 
have a 17\% branching ratio for decaying into muons, and have a large 
forward-backward asymmetry; hence, it
could cause a significant background to the asymmetry measurement. 
To reduce these backgrounds, we required additional tight selection 
criteria, which were:

\begin{itemize}

\item At least two charged tracks existed in each hemisphere.

\item A tighter cut on the invariant jet mass in the 
criteria(4): $M_{\rm jet} >$ 3.5 GeV/$c^{2}$.

\end{itemize}

A total of 29,561 events passed the above requirements. Using Monte-Carlo 
simulations, we determined the efficiency of this hadronic event 
selection to be $65.08 \pm 0.07 $\%(systematic error only). 
The remaining background in the selected events were estimated 
to be 0.2\% from $\tau$ pairs and
0.1\% from two-photon events, according to a Monte-Carlo study
\cite{BG-MC,BG-MC2}. These are negligible, compared with the statistical
errors and the other systematic errors in this analysis. 
Out of this hadronic event sample we excluded those events 
that had no high voltage on MDC. After this selection, 
27,614 hadronic events remained.

\subsection{Muon identification}

Muon identification in hadronic events suffered from fake muons, such
as punch-throughs and decay-in-flights of hadrons. In order to discriminate 
prompt muons from fake muons, following criteria were employed: 

\begin{itemize}

\item[(i)]The track had to pass the ``good'' track cuts, described in the 
previous subsection.   

\item[(ii)]The momentum of the track had to be greater 
than 2.5 GeV/c and $|\cos\theta| < 0.6$.  

After the track passed (i) and (ii), the track was  
extrapolated to MDC, assuming that it was a muon. 
We then applied the following.

\item[(iii)]All three $r-\phi$ superlayers of MDC had to have 
at least one hit in each superlayer within 20 cm 
or 3$\sigma_{track}$ from the extrapolated tracks. 
Here, $\sigma_{track}^{2}$ is the quadratic sum of the track extrapolation 
errors, the multiple Coulomb scattering errors, detector space resolution, 
and the error due to the detector alignment. 
In addition, we required that at least one superlayer had to have 
adjacent firing cells in the other layer of the same superlayer.

\item[(iv)]MDC hits were not allowed to be shared by the other tracks.

\end{itemize}

\begin{figure}
\input epsf
\begin{center}
\leavevmode
\epsfxsize=15.0cm
\epsfbox{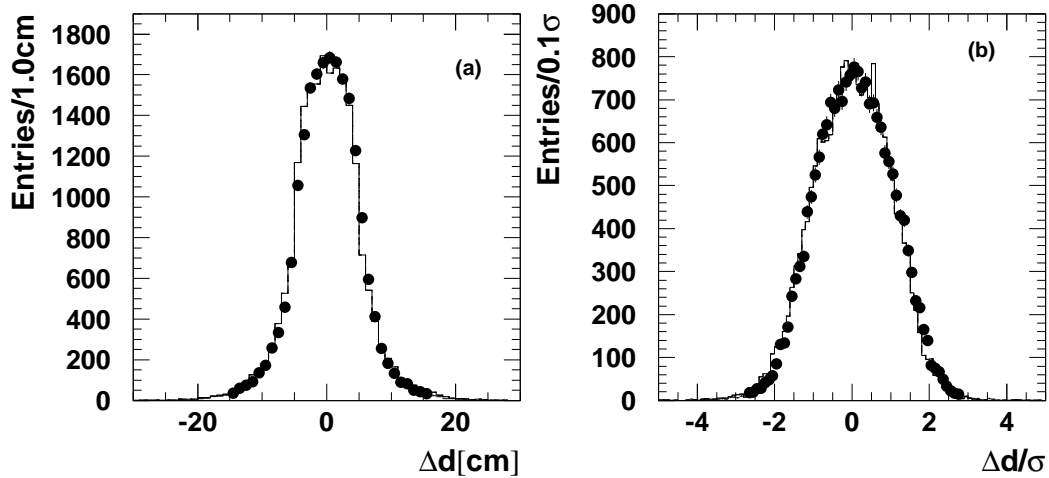}
\caption{$\Delta d$ (a) and $\Delta d/\sigma_{track}$ (b)
for muons. The dots with error bars are cosmic-ray data. Histograms are Monte
Carlo prediction.}
\label{fig:mu-ID}
\end{center}
\end{figure}

\begin{figure}
\input epsf
\begin{center}
\leavevmode
\epsfbox{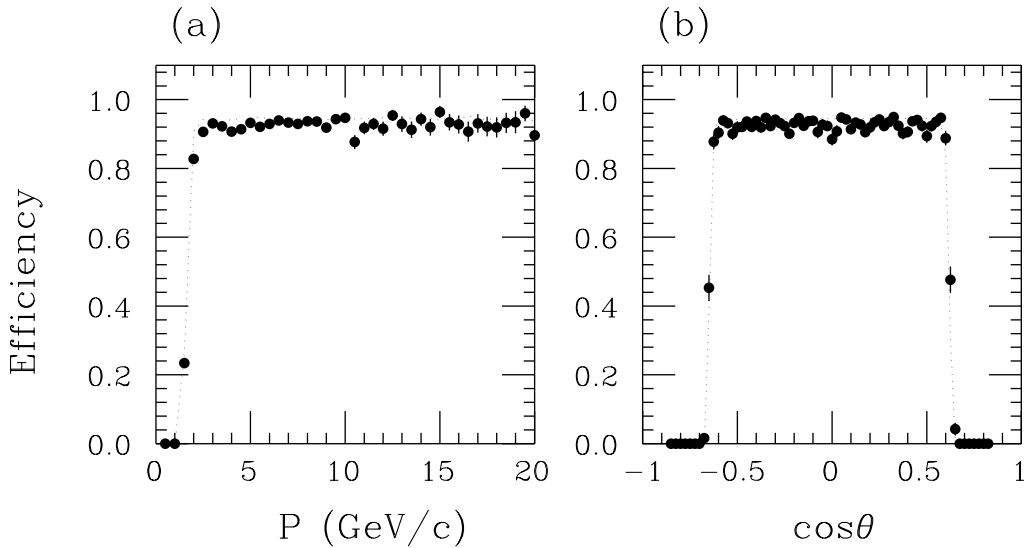}
\caption{Muon identification efficiency measured through cosmic-ray 
muons (dot with error bars) and Monte-Carlo prediction (dotted line) as a
function of the momentum (a) and polar angle (b).}
\label{fig:efficiency}
\end{center}
\end{figure}

The capability of the Monte-Carlo to simulate muons has been tested 
with cosmic-ray muons. Figure \ref{fig:mu-ID} shows the distance 
between the extrapolated track and the associated hit, $\Delta d$, 
and $\Delta d/\sigma_{track}$. As can be seen in these figures, 
the Monte-Carlo simulation and the data are in good agreement. 
The muon identification efficiencies as a 
function of momentum $p$ and also as a function of $\cos\theta$ are shown in 
Figure \ref{fig:efficiency}. The efficiency shows a plateau over 2.5 GeV/c in 
$p$ and $|\cos\theta|<0.6$. The measured efficiency for muon identification 
is 93\% at the plateau, which agrees with the Monte-Carlo 
prediction within 1\%. The inefficiency was due to dead wires 
in MDC (3\%), dead regions in the sector boundaries (3\%) and 
cuts on the distance between the MDC hits and the track (1\%).

The number of inclusive muon events was 1,328, 
after selecting the hadronic events by muon identification. 
If an event had two or more muon 
candidates, we chose the one with the highest momentum as the muon 
for tagging c and b quarks.

\section{Analysis}
\label{sec:analysis}

\subsection{Monte Carlo simulation}
\label{sec:analysis-mc}

 We used JETSET7.3\cite{L73} for $e^{+}e^{-} \rightarrow 
q\bar{q}$ event generations. The used parameters in JETSET7.3 were
tuned-up for hadronic events using the event shape 
data \cite{Ohnishi}. In the hadronization, 
we used the LUND symmetric function for the fragmentation function 
for light quarks (u,d,s) with $a = 0.413$ and $b = 0.9$. 
For heavy quarks (c,b), a function by 
Peterson {\it et al.}\cite{Peterson} was used 
with $\epsilon_{c} = 0.05$ and $\epsilon_{b} = 0.01$. 
For the standard model parameters, we used $\sin^{2}\theta_{W} = 0.2315,
M_{Z^{0}} = 91.187$ GeV/c$^{2}$, and $\Gamma_{Z^{0}} = 2.490$ GeV\cite{PDG96}. 


For a detector simulation, we used the TOPAZ detector 
simulator, which simulated the behaviors of the particles 
in the TOPAZ detector: such as the energy loss, 
multiple scattering, decay-in-flights and the detector signals.
For the simulations of particle's interactions with 
the detector material, EGS4\cite{EGS4} was used for  
electromagnetic processes and GHEISHA-7\cite{GHEI7} for nuclear 
interactions. We used 314,463 hadronic events for the studies described
in this paper.

\begin{figure}
\input epsf
\begin{center}
\leavevmode
\epsfxsize=15.0cm
\epsfbox{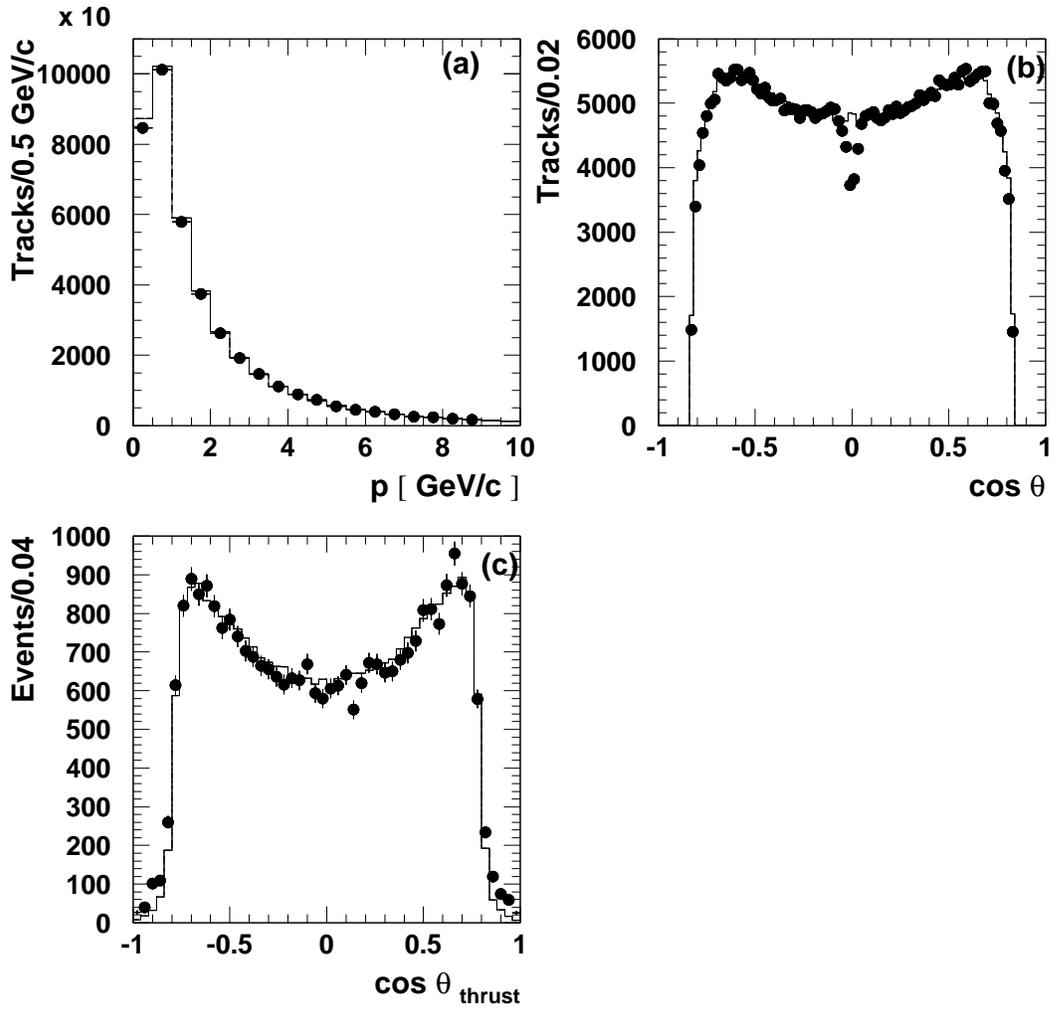}
\caption{Hadronic events; momentum (a) and
polar angle (b) of good tracks, and angular distribution of
thrust axis (c). The dots with error bars are the data; 
the histograms are Monte-Carlo results.}
\label{fig:hadron}
\end{center}
\end{figure}

To test the validity of the hadronic event selection and the Monte-Carlo 
simulation, we examined the general features of hadronic events.
Figure \ref{fig:hadron} shows the distributions of the momentum (a) 
and the polar angle (b) of the track, 
and the polar angle of the thrust axis (c). 
The data and the Monte-Carlo results are in good agreement, except for a dip 
observed at $\theta=90^{\circ}$ in the polar-angle distribution. 
This dip is due to the effect of the central membrane of TPC. 
The track reconstruction inefficiency at the membrane is 0.4\% for  
hadronic events and 0.8\% for inclusive muon events. The effect of this 
inefficiency is negligible on the cross section and asymmetry results. 

\subsection{Background estimation}

 Possible background sources to prompt muons are
hadron punch-throughs, muons from decay-in-flights of
light hadrons (mainly $\pi^{\pm}$ and $K^{\pm}$) and accidental hits
in MDC due to the beam backgrounds or cosmic rays. To estimate the
rate of accidental hits, we applied the muon identification criteria to the
electron (positron) tracks in Bhabha events, obtained in the same experimental
period. No track was identified as a muon. From this result, the accidental
hit rate was considered to be negligible.

\begin{figure}
\input epsf
\begin{center}
\leavevmode
\epsfxsize=10.0cm
\epsfbox{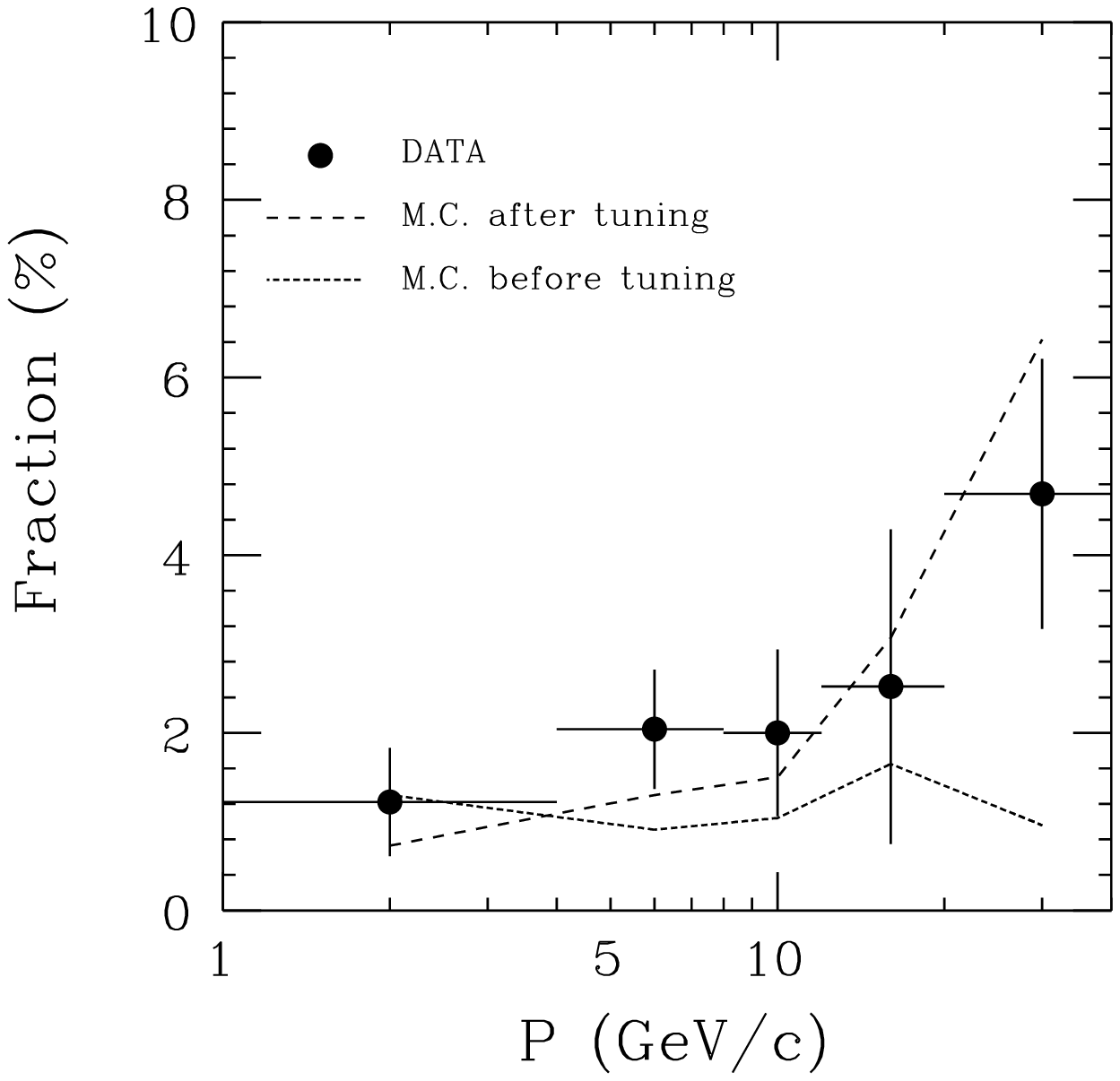}
\caption{Punch-through rate as a function of momentum derived through the pions
from $\tau$-pairs.
The dots with error bars are the data. The dotted and dashed lines are
Monte Carlo simulations before and after tuning the parameters in GHEISHA.}
\label{fig:scale-factor-mdc}
\end{center}
\end{figure}

The other background sources were evaluated using Monte-Carlo
simulations. Since the pion and kaon lifetimes and decay modes are well
established, the decay-in-flights of light hadrons can be precisely
calculated in the simulation. On the other hand,
the rate of hadron punch-throughs depends on nuclear interactions with
detector materials.
Therefore, it must be verified experimentally. For this purpose,
we studied the punch-through of pions by using 1,202 charged pions from
$\tau \rightarrow \pi\pi\pi\nu$ and $\tau \rightarrow
\rho\nu \rightarrow \pi^{\pm}\pi^{0}\nu$ in the 
$e^{+}e^{-} \rightarrow \tau^{+}\tau^{-}$ events.
Figure \ref{fig:scale-factor-mdc} shows the experimental ratio together
with the Monte-Carlo expectation as a function of the momentum.  As
can be clearly seen in this figure, it was found that the 
Monte-Carlo prediction significantly underestimated the rate, 
especially in the high-momentum region
with the default parameters.  To correct this difference, 
we tuned up the most influential
parameter, which is the total cross section of pions (kaons) interacting
with materials in the GHEISHA routine.  We introduced a correction
factor to scale the total cross section as a function of the momentum, as
listed in Table~\ref{tab:scale}. In the high-momentum region of $p >
10$ GeV/c, the correction factor was determined through the pion punch-through
rate in the $\tau$ pair events.  For $p< 3$ GeV/c, the correction factor
was estimated through energy deposits of hadrons in BCL, and it was
fitted to a function. These two correction factors were linearly
connected in  the intermediate momentum regions, i.e. $3 < p < 10$ GeV/c.

The results of the tuned-up Monte-Carlo simulation are also shown in 
Figure \ref{fig:scale-factor-mdc}. The discrepancy between the 
Monte-Carlo results and the data was reduced significantly 
with the tuned-up Monte-Carlo procedure. 

\begin{table}[h]
  \caption{Correction factor for the cross section for each momentum region.}
  \label{tab:scale}
  \begin{center}
    \begin{tabular}{ll}
      \hline \hline
	Momentum region (GeV/c)& correction factor $f(p)$\\
	\hline
	$p<3$ & $f(p) = -0.013p^{2} + 0.139p + 0.826$ \\
	$3 \leq p<10$ & $f(p) = -0.054p + 1.295$ \\
	$10\leq p$ & $f(p) = 0.726$ \\
      \hline \hline
    \end{tabular}
  \end{center}
\end{table}

\begin{figure}
\input epsf
\begin{center}
\leavevmode
\epsfxsize=15.0cm
\epsfbox{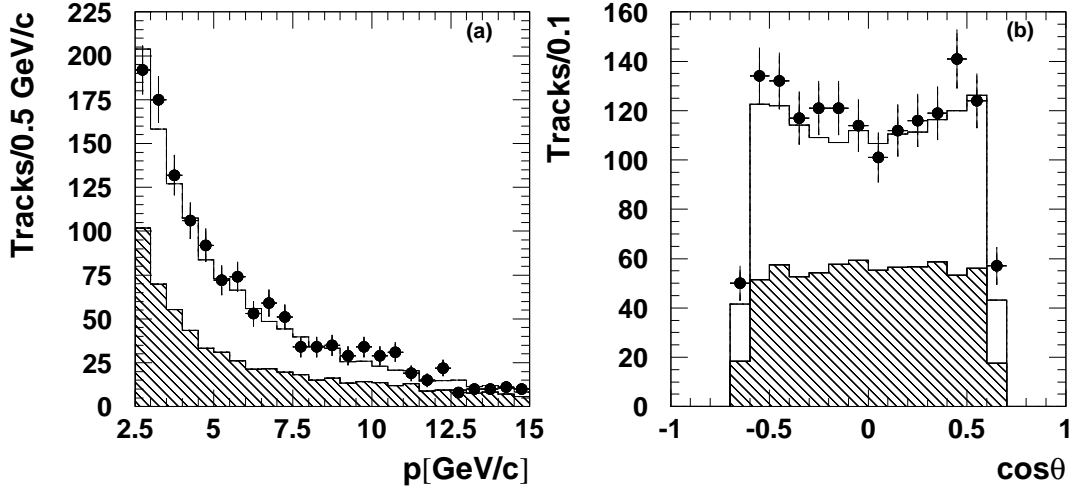}
\caption{Distributions of momentum (a) and polar angle (b) of inclusive muons. 
The dots with error bars are data and the histograms are Monte-Carlo results. 
The open regions show the contributions from c and b quarks. 
The shaded regions are the background from light quarks.}
\label{fig:inclusive-mu}
\end{center}
\end{figure}

In the inclusive muon sample, the fractions of decay-in-flight and 
hadron punch-through were estimated to be 25\% and 23\%, respectively, 
by using the tuned-up Monte-Carlo program.
Figure \ref{fig:inclusive-mu} shows the momentum spectrum (a) and polar-angle 
distribution (b). 
The background from light quarks was obtained by a Monte-Carlo 
simulation using the total number of hadronic events for normalization. 

\subsection{Flavor separation}

 Inclusive muon events can be categorized into 
four groups by the sources: (A) prompt muons from primary b-decays 
($b \rightarrow \mu$), (B) cascade muons from b $\rightarrow$ c($\bar{c}$) 
decays ($b \rightarrow c \rightarrow \mu$), (C) prompt muons from 
primary c-decays ($c \rightarrow \mu$), and (D) backgrounds from 
decay-in-flights and punch-throughs ($others$). 
In this analysis, we included the $b \rightarrow \tau \rightarrow \mu$ mode 
in category (A).
Table \ref{tab:composition} gives the results of Monte-Carlo 
studies, giving the fraction of each components in the inclusive muon events. 

\begin{figure}
\input epsf
\begin{center}
\leavevmode
\epsfxsize=15.0cm
\epsfbox{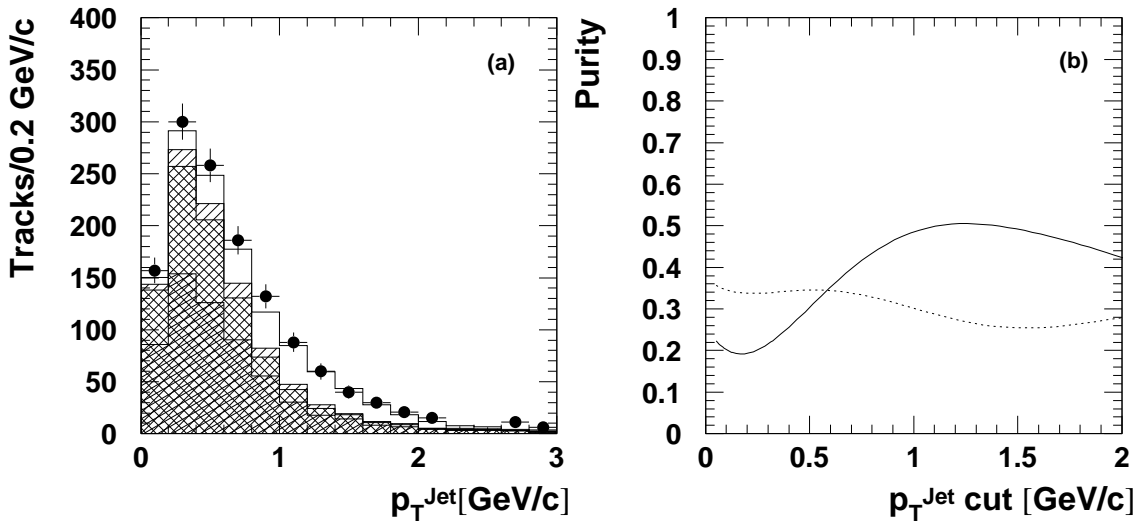}
\caption{ $p^{jet}_{T}$ distribution of inclusive muons (a) and 
purities of b and c-quarks as a function of $p^{jet}_{T}$ cut (b). 
In (a), dots with errors show the data. 
Histograms are Monte Carlo: $b \rightarrow \mu$(open), 
$b \rightarrow c \rightarrow \mu$(hatched), 
$c \rightarrow \mu$(double-hatched) and 
$others$(triple-hatched), respectively.
In (b), solid curve shows the b-quark purity 
and dotted curve shows the c-quark purity.}
\label{fig:ptcut}
\end{center}
\end{figure}

We used the thrust axis to determine the primary quark direction. 
The accuracy of the quark direction determined by this method was estimated 
to be $5^{\circ}$, based on a study by a Monte-Carlo simulation. 
We define the angle of the quark direction by
$-Q\cos\theta_{trust}$, where $Q$ is the charge of the tagged muon 
and $\theta$ is the angle of the thrust axis with respect to the beam axis. 
To separate the quark flavors, we used the muon transverse momentum 
with respect to the jet axis, $p_{T}^{jet}$. Jets were reconstructed 
using the JADE jet clustering algorithm\cite{JET} with a scaled 
invariant mass cut, $Y_{cut}(= M_{ij}/E_{vis}) = 0.04$. 
Muons from b-quarks have a larger $p_{T}^{jet}$ compared with those 
from c-quarks, due to the heavy b-quark mass.
Using the $p_{T}^{jet}$ cut at 0.8 GeV/c, we classified the inclusive 
muon events into a b-enriched sample ($p_{T}^{jet} \ge 0.8$ GeV/c) and 
a c-enriched sample ($p_{T}^{jet}<0.8$ GeV/c). 
We chose $p_{T}^{jet} = 0.8$ GeV/c, because the systematic 
uncertainty was the smallest at 0.8 GeV/c. 
The purity of the c- and b-quarks as a function of the $p_{T}$ cut 
was studied by a Monte-Carlo simulation. 
Figure \ref{fig:ptcut} shows the b, c-quark purities as a
function of the $p_{T}$ cut. For b-quarks, the purity was calculated by 
summing up all of the b-quarks with $p_{T}$ greater than a given $p_{T}$ cut, 
divided by the total number of inclusive muon tracks 
in the same $p_{T}$ region. 
For c-quarks, summing was done for $p_{T}$ less than 
a given $p_{T}$ cut.

\newcommand{\lw}[1]{\smash{\lower2.0ex\hbox{#1}}}
\begin{table}[h]
  \caption{Percentage of muons from each source in the inclusive muon sample.}
  \label{tab:composition}
  \begin{center}
    \begin{tabular}{llll}
      \hline \hline
      \lw{Source} & \multicolumn{3}{c}{\rm Fraction$(\%)$} \\
             & $p^{jet}_{T} < 0.8$ GeV/c & $p^{jet}_{T} \geq 0.8$ GeV/c & total \\
      \hline
      $b \rightarrow \mu$ & 9.7 & 41.9 & 20.1 \\
      $b \rightarrow c \rightarrow \mu$ & 5.8 & 5.2 & 5.6 \\
      $c \rightarrow \mu$ & 31.6 & 13.2 & 25.7 \\
      $others$ & 52.8 & 39.6 & 48.5 \\
      \hline \hline
    \end{tabular}
  \end{center}
\end{table} 

\subsection{Simultaneous fit for b and c-quark}

 In a previous paper\cite{shimonaka}, we derived only the b-quark parameters
by assuming the standard-model parameters for the c-quark because of low
statistics. This time, we 
performed a four-parameter fit of ($R_{b\bar{b}}, R_{c\bar{c}},
A_{FB}^{b}, A_{FB}^{c}$) to $p$ and $-Q\cos\theta_{thrust}$ 
($p_{T}^{jet} \ge 0.8$ GeV/c and $<0.8$ GeV/c) distributions. 
Definitions of $R_{q\bar{q}}$, and $A_{FB}$ are given by
\begin{equation}
R_{q\bar{q}}\equiv\sigma(e^{+}e^{-}\rightarrow q\bar{q})/
\sigma(e^{+}e^{-}\rightarrow hadrons)
\end{equation}
and
\begin{equation}
A_{FB}\equiv{\sigma_{F}-\sigma_{B}\over\sigma_{F}+\sigma_{B}}
\end{equation}
where $\sigma$ represents the lowest order (SU(3) $\times$ SU(2)$_{L}$
$\times$ U(1)) cross-section and 
$\sigma_{F}(\sigma_{B})$ is the cross-section for quark to
travel forward (backward) with respect to the $e^{-}$ direction.
The $\chi^{2}$ of the fit was defined by
\begin{eqnarray}
\chi^{2}_{total} & = & \chi^{2}_{\cos\theta(p_{T}^{jet} \ge 0.8 GeV/c)}
		 + \chi^{2}_{\cos\theta(p_{T}^{jet}<0.8 GeV/c)}
		 + \chi^{2}_{p}\nonumber\\
		 & = & \sum_{i=1}^{8}\frac{(N^{\mu}_{i} - 
\tilde{N}^{\mu}_{i})^{2}}{\sigma_{N^{\mu}_{i}}^{2}}
		 + \sum_{j=1}^{8}\frac{(N^{\mu}_{j} - 
\tilde{N}^{\mu}_{j})^{2}}{\sigma_{N^{\mu}_{j}}^{2}}
		 + \sum_{k=1}^{6}\frac{(N^{\mu}_{k} - 
\tilde{N}^{\mu}_{k})^{2}} {\sigma_{N^{\mu}_{k}}^{2}},
\end{eqnarray}
where $N^{\mu}_{i}$ and $\sigma_{N^{\mu}_{i}}$ are the number of inclusive 
muon events and the statistical error for each bin, respectively. 
$\tilde{N}^{\mu}_{i}$ is 
the number of expected events for each bin, which is defined by 
\begin{eqnarray}
	\tilde{N}^{\mu}_{i} & = & N^{exp}_{had}\{R_{b\bar{b}}\cdot
2Br^{\rm sum}(b \rightarrow \mu)F(A_{FB}^{b})_{i}\cdot 
C^{b \rightarrow \mu}_{i} \nonumber\\
	& + & R_{b\bar{b}}\cdot2Br(c \rightarrow \mu)
(1+\alpha)F(-\frac{1-\alpha}{1+\alpha}A_{FB}^{b})_{i}\cdot C^{b \rightarrow c
\rightarrow \mu}_{i} \nonumber\\
	& + & R_{c\bar{c}}\cdot2Br(c \rightarrow
\mu)F(-A_{FB}^{c})_{i}\cdot C^{c \rightarrow \mu}_{i}\} \nonumber\\
	& + & N^{others}_{i}
\end{eqnarray}
for $-Q\cos\theta_{thrust}$ bins, and 
\begin{eqnarray}
	\tilde{N}^{\mu}_{k} & = & N^{exp}_{had}\{R_{b\bar{b}}\cdot
2Br^{\rm sum}(b \rightarrow \mu)\cdot W^{b \rightarrow \mu}_{k} \nonumber\\ 
& + & R_{b\bar{b}}\cdot2Br(c \rightarrow \mu)(1+\alpha)\cdot 
W^{b \rightarrow c \rightarrow \mu}_{k} \nonumber\\
	& + & R_{c\bar{c}}\cdot2Br(c \rightarrow \mu)\cdot 
W^{c \rightarrow \mu}_{k}\} \nonumber\\
	& + & N^{others}_{k}
\end{eqnarray}
for $p$ bins, where $N^{exp}_{had}$ is the total number of 
hadronic events. 
$Br^{\rm sum}(b \rightarrow \mu)$ is the combined branching ratio of
$Br(b \rightarrow \mu)$ and $Br(b \rightarrow \tau \rightarrow \mu)$.
$Br(b \rightarrow \mu)$ and $Br(c \rightarrow \mu)$ 
are the branching ratios of $b \rightarrow \mu$ and $c \rightarrow \mu$,
which are equivalent to the branching ratios for muons semileptonically decayed 
from b and c hadrons, respectively. 
$Br(b \rightarrow c\bar{c}s)$ is denoted as $\alpha$
in the above equations. 
The used branching ratios are listed in Table \ref{tab:br}.
$F(A_{FB}^{q})_{i}$ is the polar-angle distribution function, integrated
in each $\cos \theta$ bin ($i$-th bin), which is written as
\begin{equation}
        F(A_{FB}^{q})_{i} = \int_i \frac{3}{8}(1 + \cos^{2}\theta +
\frac{8}{3}A_{FB}^{q}\cos\theta) d\cos\theta.
\end{equation}
\begin{table}[h]
  \caption{Branching ratios used in the fits.}
  \label{tab:br}
  \begin{center}
    \begin{tabular}{|r|r|}
      \hline \hline
	Mode & Branching ratio(\%) \\
      \hline
      $Br^{\rm sum}(b \rightarrow \mu)$		& $11.3 \pm 0.5$ \\
      $Br(b \rightarrow \mu)$\cite{PDG98}	& $10.8 \pm 0.5$ \\
      $Br(b \rightarrow \tau \rightarrow \mu)$\cite{b-tau} & $0.45$ \\
      $Br(c \rightarrow \mu)$\cite{C-mu}	& $9.0 \pm 0.7$ \\
      $Br(b \rightarrow c\overline{c}s)$\cite{CCS} & $21.9 \pm 3.7$ \\
      \hline \hline
    \end{tabular}
  \end{center}
\end{table} 
$C_{i}^{mode}$ is the correction factor for the each $\cos \theta$ bin
from the {\it decay mode}, given by 
\begin{eqnarray}
C^{mode}_{i} =
\frac{(1+\delta)^{q\bar{q}}_{i}}{(1+\delta)^{had}_{total}}\frac{\eta^{q\bar{q}}
_{i}}{\eta^{had}}
\varepsilon^{\mu}_{i}, 
\end{eqnarray}
where $(1+\delta)_{i}^{q\bar{q}}$ is the combined correction factor of the 
QED (initial state photon radiation) and QCD (parton shower) radiative 
corrections for the $i$-th bin of $e^{+}e^{-} \rightarrow q\bar{q}$.
Similarly, $(1+\delta)_{total}^{had}$ is a radiative correction 
factor for the total hadronic cross section.
These correction factors were estimated through a Monte-Carlo 
simulation based on the LUND event generator, JETSET7.3. 
$\eta_{i}^{q\bar{q}}$ is the acceptance correction for the 
$i$-th bin of each mode, $\eta^{had}$ is that for hadronic events, 
and $\varepsilon^{\mu}_{i}$ is the muon identification efficiency 
for the $i$-th bin ($\cos \theta$ bin). $W^{mode}_{k}$ is the correction
factor for each $p$ bin, which is given by
\begin{eqnarray}
W^{mode}_{k} =
\frac{(1+\delta)^{q\bar{q}}_{k}}{(1+\delta)^{had}_{total}}
\frac{\eta^{q\bar{q}}_{k,mode}}{\eta^{had}}\varepsilon^{\mu}_{k}.
\end{eqnarray}
Table \ref{tab:num-corr} gives the correction factors 
for each bin. $N_{i}^{others}$ is the number of background events 
for the $i$-th bin, which is also listed in Table \ref{tab:num-corr}. 
The numbers were derived through Monte Carlo simulations.

\begin{figure}
\input epsf
\begin{center}
\leavevmode
\epsfxsize=15.0cm
\epsfbox{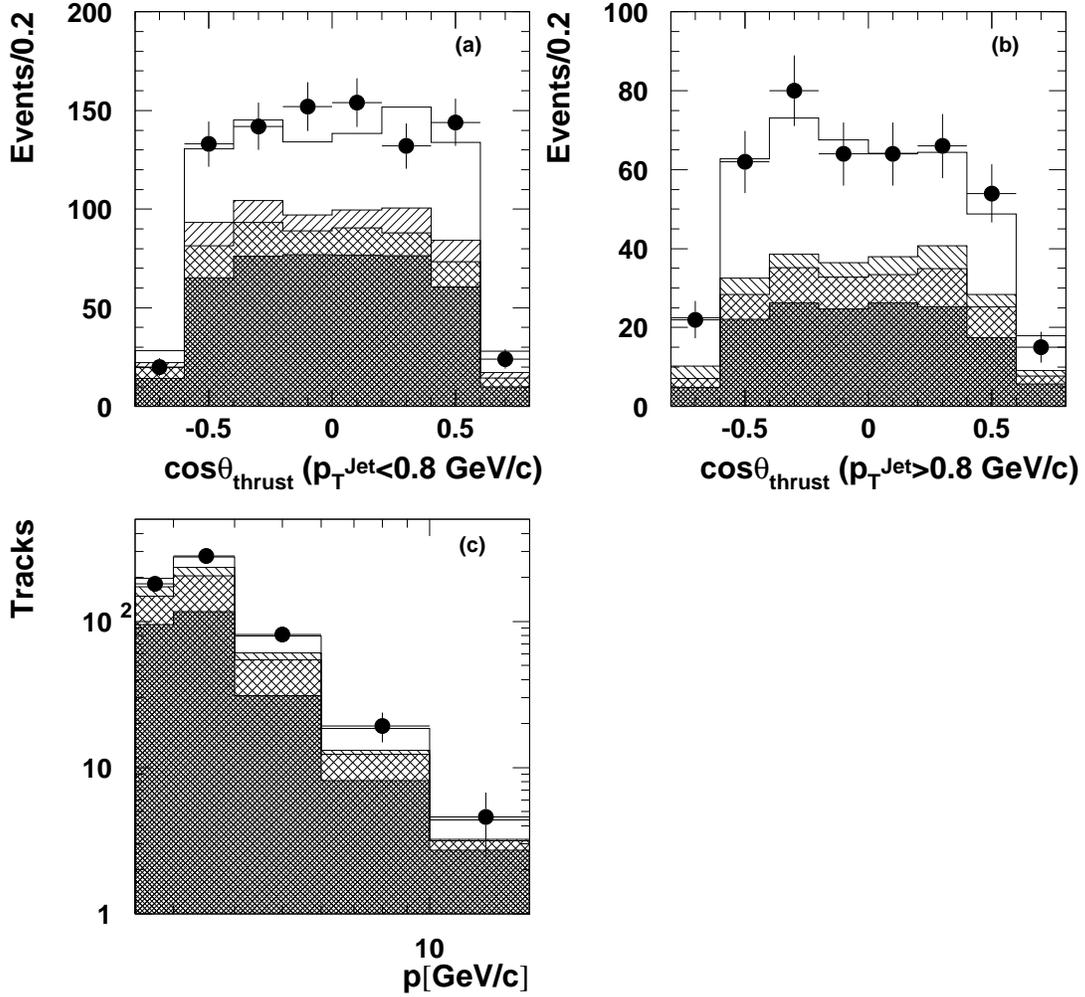}
\caption{Results of the four-parameter fit: 
$\cos\theta_{thrust}$ for the c-enriched sample($p_{T}^{jet}<0.8$ GeV/c) (a)
and the b-enriched sample($p_{T}^{jet}\ge0.8$ GeV/c) (b), and 
momentum distributions (c). The dots with error bars are the data. 
The histograms are the results from the four-parameter fit: 
$b \rightarrow \mu$(open), $b \rightarrow c \rightarrow \mu$(hatched), 
$c \rightarrow \mu$(double-hatched) and 
$others$(triple-hatched), respectively.}
\label{fig:flavor-fit}
\end{center}
\end{figure}

\begin{figure}
\input epsf
\begin{center}
\leavevmode
\epsfbox{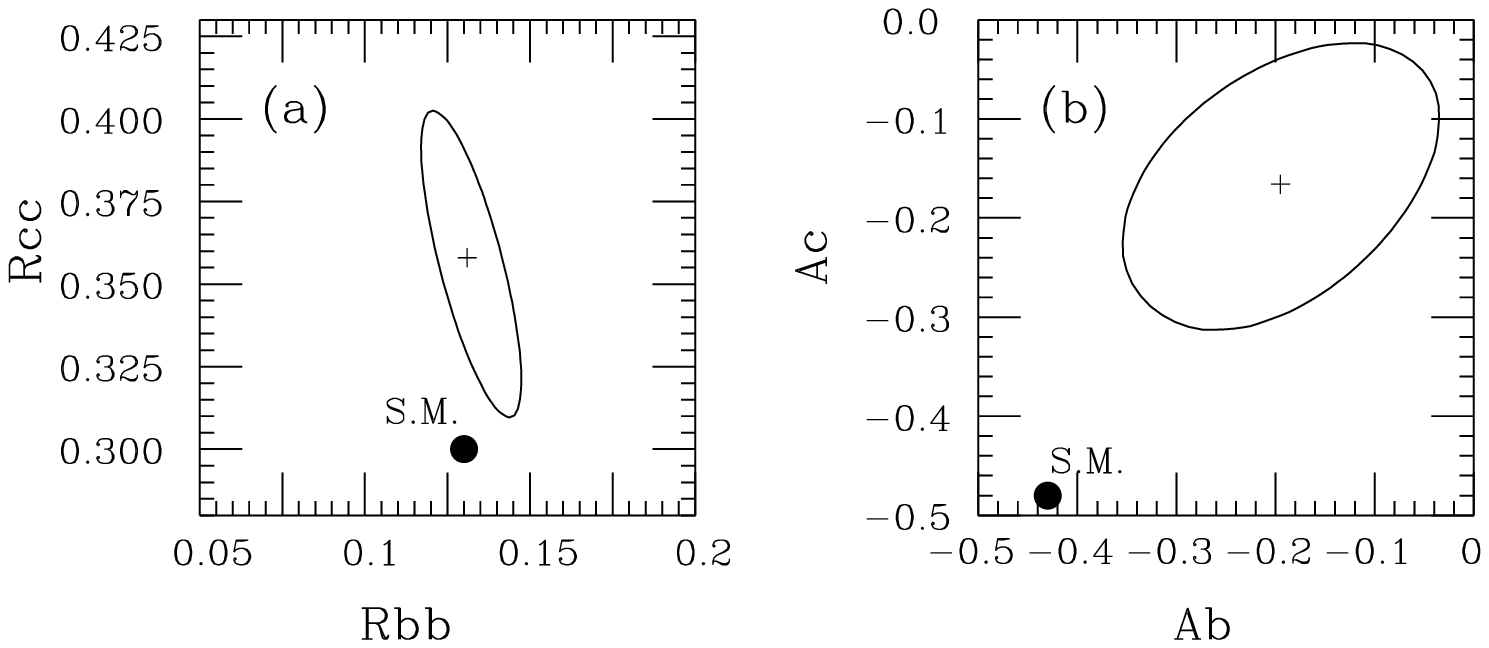}
\caption{Results of the four-parameter fit for $R_{q\bar{q}}$ (a) 
and $A_{FB}^{q}$ (b).
The cross bar indicates the central value and the ellipse shows 1$\sigma$ 
contour. The errors are only statistical. The points show the 
standard-model prediction.}
\label{fig:cont}
\end{center}
\end{figure}

As a result of the four-parameter fit (Figure \ref{fig:flavor-fit} and 
\ref{fig:cont}), we obtained 
$R_{b\bar{b}} = 0.13 \pm 0.02$, $R_{c\bar{c}} = 0.36 \pm 0.05$,
$A_{FB}^{b} = -0.20 \pm 0.16$ and $A_{FB}^{c} = -0.17 \pm 0.14$, 
with $\chi^{2}/\mbox{D.O.F} = 14.93/18$. The errors are statistical only.
The correlation coefficients obtained from the fit is given in Table
\ref{tab:coefficients}. The 1$\sigma$ contours of the fit for $R_{cc}$
v.s. $R_{bb}$ and $A_{c}$ v.s. $A_{b}$ are shown in Figure \ref{fig:cont}.

\begin{table}[h]
  \caption{Number of events and correction factors in each bin.}
  \label{tab:num-corr}
  \begin{center}
   (a) \hspace{0.5cm}Correction factor for each $\cos \theta$ bin: 
$p_{T}^{Jet} < 0.8$ (GeV/c)\\
   \begin{tabular}{cccccc}
      \hline \hline
      $\cos\theta$ bin & $N^{\mu}_{i}$ & $N^{others}_{i}$ 
	&{\rm$C^{b \rightarrow\mu}_{i}$}
	&{\rm$C^{b \rightarrow b \rightarrow \mu}_{i}$}
	&{\rm$C^{c \rightarrow \mu}_{i}$}\\
      \hline
      -0.8 $\sim$ -0.6 &  20.$\pm$ 4.5& 14.0$\pm$1.1& 0.049& 0.034& 0.038\\
      -0.6 $\sim$ -0.4 & 133.$\pm$11.5& 65.0$\pm$2.4& 0.178& 0.180& 0.271\\
      -0.4 $\sim$ -0.2 & 142.$\pm$11.9& 76.2$\pm$2.6& 0.222& 0.187& 0.319\\
      -0.2 $\sim$  0.0 & 152.$\pm$12.3& 76.9$\pm$2.6& 0.183& 0.137& 0.288\\
       0.0 $\sim$  0.2 & 154.$\pm$12.4& 76.6$\pm$2.6& 0.234& 0.149& 0.273\\
       0.2 $\sim$  0.4 & 132.$\pm$11.5& 76.5$\pm$2.6& 0.201& 0.176& 0.313\\
       0.4 $\sim$  0.6 & 144.$\pm$12.0& 60.5$\pm$2.3& 0.209& 0.128& 0.253\\
       0.6 $\sim$  0.8 &  24.$\pm$ 4.9&  9.8$\pm$0.9& 0.064& 0.029& 0.044\\
      \hline \hline
    \end{tabular}
\\
   \vspace*{1cm}
   (b) \hspace{0.5cm}Correction factor for each $\cos \theta$ bin: 
$p_{T}^{Jet} \geq 0.8$ (GeV/c)\\
   \begin{tabular}{cccccc}
      \hline \hline
      $\cos\theta$ bin & $N^{\mu}_{i}$ & $N^{others}_{i}$ 
	&{\rm$C^{b \rightarrow\mu}_{i}$}
        &{\rm$C^{b \rightarrow c \rightarrow \mu}_{i}$}
	&{\rm$C^{c \rightarrow \mu}_{i}$} \\
      \hline
      -0.8 $\sim$ -0.6 & 22.$\pm$4.7&  4.8$\pm$0.7 & 0.106& 0.042& 0.014\\
      -0.6 $\sim$ -0.4 & 62.$\pm$7.9& 22.0$\pm$1.4 & 0.326& 0.062& 0.046\\
      -0.4 $\sim$ -0.2 & 80.$\pm$8.9& 26.3$\pm$1.5 & 0.449& 0.057& 0.070\\
      -0.2 $\sim$  0.0 & 64.$\pm$8.0& 24.8$\pm$1.5 & 0.476& 0.063& 0.062\\
       0.0 $\sim$  0.2 & 64.$\pm$8.0& 26.3$\pm$1.5 & 0.443& 0.073& 0.050\\
       0.2 $\sim$  0.4 & 66.$\pm$8.1& 25.2$\pm$1.5 & 0.411& 0.080& 0.060\\
       0.4 $\sim$  0.6 & 54.$\pm$7.3& 17.4$\pm$1.2 & 0.334& 0.037& 0.040\\
       0.6 $\sim$  0.8 & 15.$\pm$3.9&  5.7$\pm$0.7 & 0.125& 0.014& 0.008\\
      \hline \hline
    \end{tabular}
\\
   \vspace*{1cm}
   (c) \hspace{0.5cm}Correction factor for each momentum bin \\
    \begin{tabular}{cccccc}
      \hline \hline
      Momentum bin (GeV/c) & $N^{\mu}_{k}$ & $N^{others}_{k}$ 
	&{\rm $W^{b \rightarrow \mu}_{k}$}
	&{\rm $W^{b \rightarrow c \rightarrow \mu}_{k}$}
	&{\rm $W^{c \rightarrow \mu}_{k}$} \\
      \hline
      2.5 $\sim$3.0 & 181.$\pm$ 13.5 & 95.0 $\pm$2.9 & 3.048 & 2.902 & 2.998\\
      3.0 $\sim$4.0 & 281.$\pm$ 16.7 & 116.9$\pm$3.2 & 4.921 & 3.764 & 4.903\\
      4.0 $\sim$6.0 & 325.$\pm$ 18.0 & 124.6$\pm$3.3 & 4.572 & 1.573 & 2.627\\
      6.0 $\sim$10.0& 309.$\pm$ 17.6 & 130.5$\pm$3.4 & 2.600 & 0.402 & 0.936\\
      10.0$\sim$16.0& 165.$\pm$ 12.8 & 98.2 $\pm$2.9 & 0.865 & 0.045 & 0.145\\
      16.0$\sim$30.0& 68. $\pm$ 8.2  & 43.2 $\pm$1.9 & 0.091 & 0.007 & 0.017\\
      \hline \hline
    \end{tabular}
  \end{center}
\end{table}

\begin{table}[h]
  \caption{The correlation coefficients for the parameters in the 
four-parameter fit.}
  \label{tab:coefficients}
  \begin{center}
    \begin{tabular}{l|llll}
      \hline \hline
             & $R_{b\bar{b}}$ & $R_{c\bar{c}}$ &
               $A_{FB}^{b}$ & $A_{FB}^{c}$ \\
      \hline
      $R_{b\bar{b}}$ & 1 & -0.77 &  0.01 & -0.17 \\
      $R_{c\bar{c}}$ &   &     1 & -0.37 &  0.16 \\
      $A_{FB}^{b}$   &   &       &     1 &  0.44 \\
      $A_{FB}^{c}$   &   &       &       &     1 \\
      \hline
      \hline
    \end{tabular}
  \end{center}
\end{table}

\subsection{Systematic errors for $R_{q\bar{q}}$ and $A^{q}_{FB}$}

\hspace*{12pt}Possible sources of systematic errors are listed in 
Table \ref{tab:systematic}. The largest systematic error comes from 
the uncertainty in the probabilities of mis-identified hadrons as muons. 
This is caused by the light quark background in the heavy 
quark samples. In order to estimate the effect of hadron mis-identifications, 
we performed the same analysis by changing the correction factor of the 
effective pion-nucleus cross section by $\pm 1 \sigma$ of the measured error 
of fake rate. The effect of the cuts on 
muon identification was studied by varying the cut values 
on momentum $p$, $\Delta d$, and $\Delta d/\sigma_{track}$ by 
$\pm 0.5$ GeV/c, $\pm 5$ cm, and $\pm 0.5 \sigma_{track}$, respectively,
and examined the change in the obtained cross sections and asymmetries. 
We studied the effect of the $p_{T}^{jet}$ cut by shifting the 
cut value by $\pm 0.1$ GeV/c. The effect of the uncertainly in the MDC 
detector acceptance was checked by varying the $|\cos\theta_{track}|$ cut 
from 0.6 to 0.58. Based on these studies, we selected the cut values that 
minimize the systematic errors. The error due to the semileptonic branching 
ratios for b and c hadrons were obtained by changing the branching
ratios by $\pm 1 \sigma$ of the quoted number. 
Changes of fragmentation parameters cause the changes in the muon 
identification efficiency and flavor separation. The error due to this effect 
was estimated by changing the fragmentation parameters by 
$\pm 1 \sigma$ of the quoted number. 
\begin{table}[h]
  \caption{Summary of systematic errors}
  \label{tab:systematic}
  \begin{center}
    \begin{tabular}{lllll}
      \hline \hline
      Source & $\Delta R_{b\bar{b}}/R_{b\bar{b}}$ & 
               $\Delta R_{c\bar{c}}/R_{c\bar{c}}$ &
               $\Delta A_{FB}^{b}/A_{FB}^{b}$ & 
               $\Delta A_{FB}^{c}/A_{FB}^{c}$ \\
      \hline
      hadron mis-ID & 4.9\%& 12.1\%& 3.8\%& 13.3\%\\
      muon ID       & $<$0.1\%& 0.6\%& 1.5\%& 0.9\%\\
      $p^{jet}_{T}$ cut & 0.9\%& 0.6\%& 1.5\%& 0.2\%\\
      MDC acceptance & $<$0.1\%& $<$0.1\%& 0.5\%& 0.9\%\\
      branching ratio & 2.8\%& 5.0\%& 2.1\%& 1.2\%\\
      fragmentation parameter & 1.7\%& 3.8\%& $<$0.1\%& $<$0.1\%\\
      \hline
      total      & 6.1\%& 13.7\%& 4.9\%& 13.8\%\\
      \hline \hline
    \end{tabular}
  \end{center}
\end{table}

\clearpage
\section{Discussion}
\label{sec:discussion}

The results of the four-parameter fits with systematic errors are
$R_{b\bar{b}} = 0.13\pm0.02(stat)\pm0.01(syst)$, 
$R_{c\bar{c}} = 0.36\pm0.05(stat)\pm0.05(syst)$, 
$A_{FB}^{b} = -0.20\pm0.16(stat)\pm0.01(syst)$ 
and $A_{FB}^{c} = -0.17\pm0.14(stat)\pm0.02(syst)$. 
Four-parameter fits of the c- and b-quark have correlations in the
parameters of the c-quark and b-quark. 
For example, an increase of $R_{c\bar{c}}$ would cause 
$R_{b\bar{b}}$ to decrease. Also, an increase of
$A_{FB}^{c}$ would cause $A_{FB}^{b}$ to increase.
To estimate this effect, we applied one quark parameter fit, by
fixing the other quark parameters to the standard-model predictions.
The used values of the standard model parameters are $R_{b\bar{b}} = 0.13$, 
$A_{FB}^{b} = -0.43$ for the c-quark fit and $R_{c\bar{c}} = 0.30$,
$A_{FB}^{c} = -0.48$ for the b-quark fit. For $A_{FB}^{b}$, 
correction for $B^{0}_{d,s}\bar{B^{0}_{d,s}}$ mixing is included using
the mixing parameter at high energy, $\chi_{B} = 12\%$.
The results of the fits are $R_{b\bar{b}} = 0.15\pm0.01$, 
$A_{FB}^{b} = -0.29\pm0.13$(b-quark fit) and 
$R_{c\bar{c}} = 0.36\pm0.03$,  $A_{FB}^{c} = -0.26\pm0.13$ (c-quark fit).
The results are summarized in Table \ref{tab:1quark}. They are consistent with 
the four-parameter fit.

\begin{table}[h]
  \caption{Summary of $R_{q\bar{q}}$ and $A_{FB}^{q}$.}
  \label{tab:1quark}
  \begin{center}
    \begin{tabular}{l|llll}
      \hline \hline
	fitting method & flavor & $R_{q\bar{q}}$ & $A_{FB}^{q}$ & $\chi^{2}$/D.O.F \\
	\hline
	\lw{four-parameter fit} & c-quark & $0.36\pm0.05$ & $-0.17\pm0.14$ & \lw{14.93/18} \\
	& b-quark & $0.13\pm0.02$ & $-0.20\pm0.16$ & \\
	\hline
	\lw{1 quark parameter fit} & c-quark & $0.36\pm0.03$ & $-0.26\pm0.13$ & 19.29/20 \\
	& b-quark & $0.15\pm0.01$ & $-0.29\pm0.13$ & 19.29/20 \\
	\hline
	\lw{Standard Model predictions} & c-quark & 0.30 & -0.48 & \\
	& b-quark & 0.13 & -0.43 (mixing $\chi_{B} = 12\%$)& \\
      \hline \hline
 	\multicolumn{5}{l}{\small The errors are statistical only.}\\
    \end{tabular}
  \end{center}
\end{table} 

\begin{figure}
\input epsf
\begin{center}
\leavevmode
\epsfxsize=15.0cm
\epsfbox{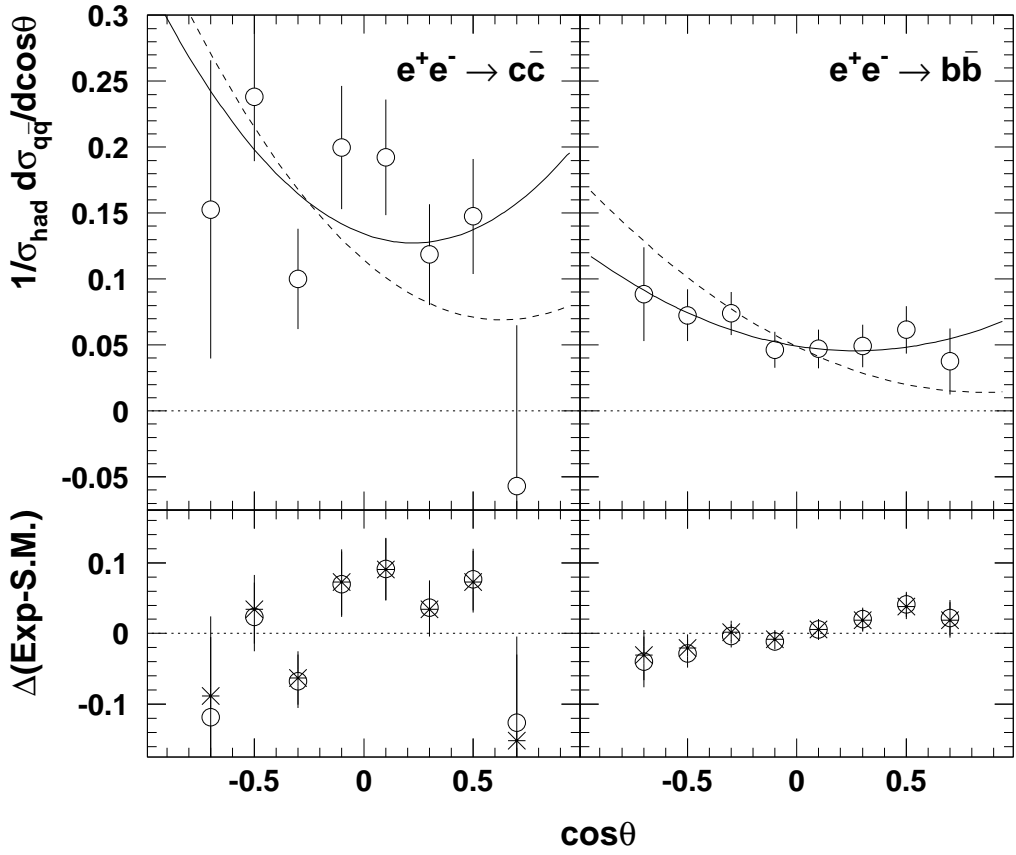}
\caption{Differential cross section of $e^{+}e^{-} \rightarrow b\bar{b}$ and 
$e^{+}e^{-} \rightarrow c\bar{c}$ (upper figures). The solid lines are the 
best fits and the dashed lines are the standard model predictions. Lower 
figures show the difference between the observed cross section and the standard
model prediction. The open circles are the result of the 
four-parameter fit and the cross marks are the result of the
1-quark parameter fit.}
\label{fig:cos-clo}
\end{center}
\end{figure}

The differential cross sections for b-quark production 
($d\sigma_{b\bar{b}}/d\cos\theta$) and c-quark production 
($d\sigma_{c\bar{c}}/d\cos\theta$) were obtained from 
the $\cos\theta_{thrust}$ distribution. In order to 
derive $d\sigma_{b\bar{b}}/d\cos\theta$, the contributions 
from light quarks and $c\bar{c}$ were subtracted, and $b\bar{b}$ was 
subtracted for $d\sigma_{c\bar{c}}/d\cos\theta$.  The results are given in 
Table \ref{tab:cross-section} and Figure \ref{fig:cos-clo}. The measured 
asymmetry for the c- and b-quarks are smaller than the standard-model 
prediction by 2.2$\sigma$ and 1.5$\sigma$, respectively. 

\begin{figure}
\input epsf
\begin{center}
\leavevmode
\epsfxsize=15.0cm
\epsfbox{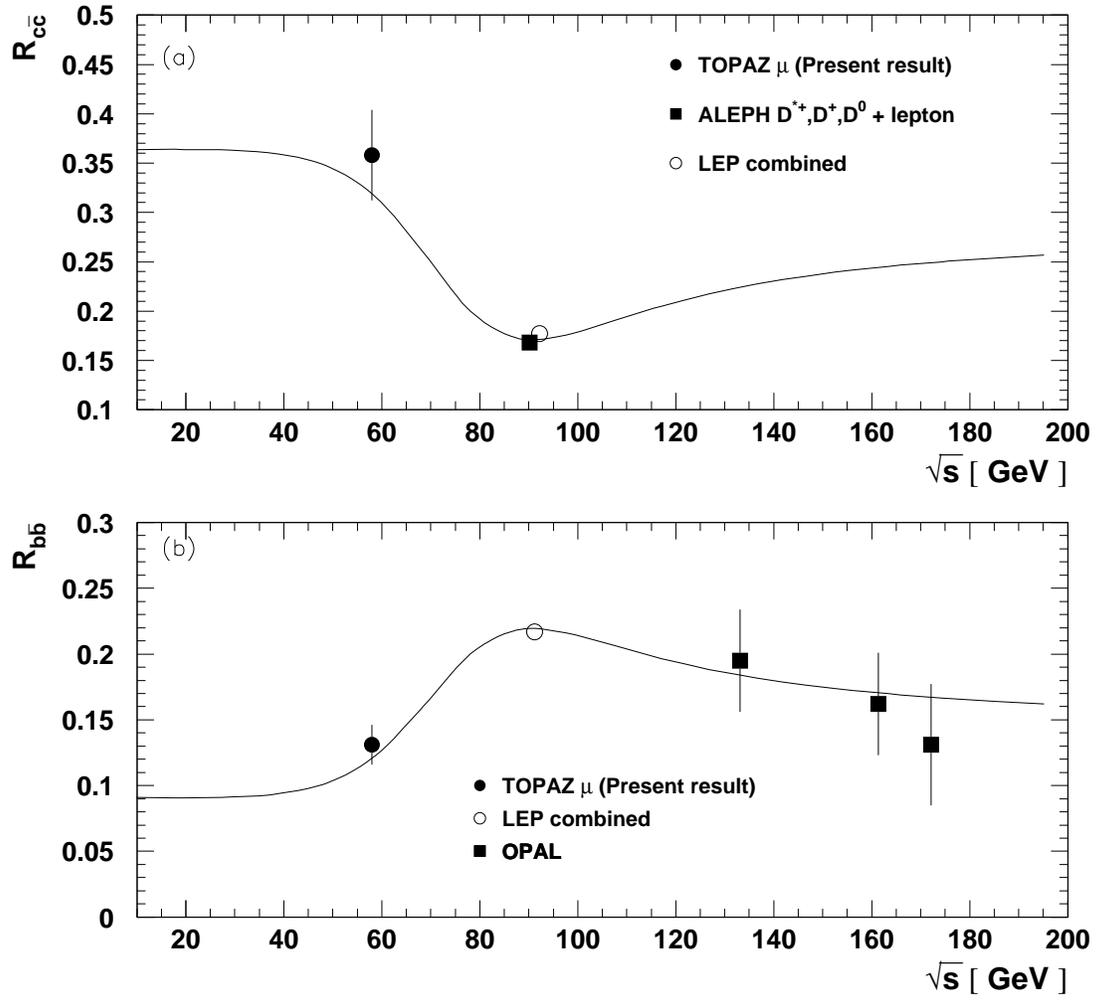}
\caption{$R_{c\bar{c}}$ (a) and $R_{b\bar{b}}$ (b) as a function of the 
center-of-mass energy.}
\label{fig:rclo}
\end{center}
\end{figure}

\begin{figure}
\input epsf
\begin{center}
\leavevmode
\epsfxsize=15.0cm
\epsfbox{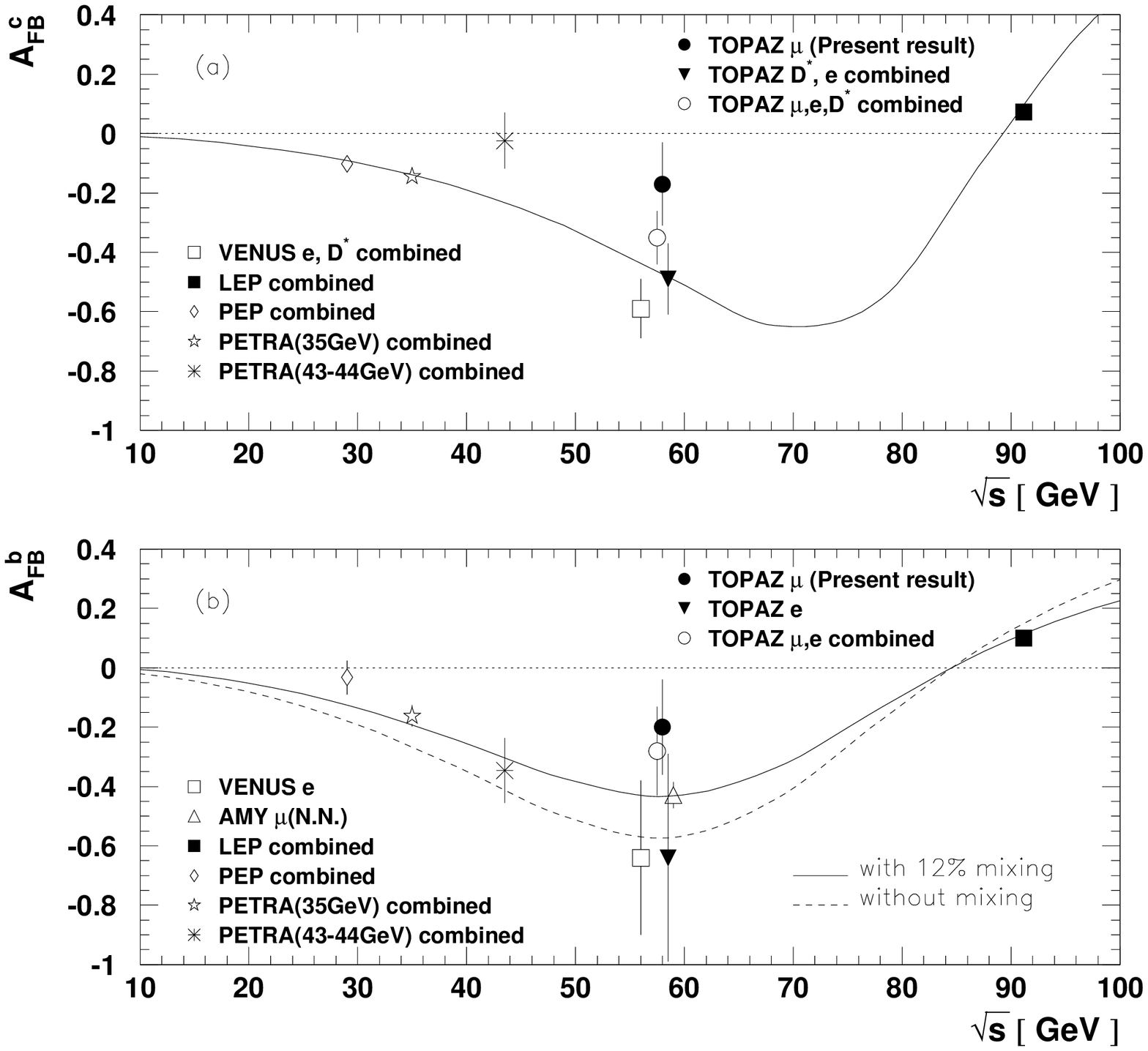}
\caption{Forward-backward charge asymmetry for c-quark (a) and b-quark 
(b) as a function of the center-of-mass energy. }
\label{fig:afb}
\end{center}
\end{figure}

Figure \ref{fig:rclo} shows the measured $R_{c\bar{c}}$ and $R_{b\bar{b}}$  
together with the other experimental data\cite{PDG98,LEP-R90,OPAL-R130}. 
The measured asymmetries
of b and c-quark are shown in Figure \ref{fig:afb} with the other 
experimental data\cite{nakano,nakano2,PDG98,AMY,VENUS,VENUS2,
PEP,PEP2,PEP3,PEP4,PEP5,PEP6,PEP7,
PETRA,PETRA2,PETRA3,PETRA4,PETRA5,PETRA6,PETRA7,PETRA8,PETRA9}. 
The asymmetries measured 
at PEP\cite{PEP,PEP2,PEP3,PEP4,PEP5,PEP6,PEP7}, 
PETRA\cite{PETRA,PETRA2,PETRA3,PETRA4,PETRA5,PETRA6,PETRA7,PETRA8,PETRA9}, 
and LEP\cite{PDG98} are combined for 
each experiment. 
We combined the present results of $A_{FB}^{q}$ with our previous measurement 
by inclusive electron ( for c and b quark ) and $D^{*\pm}$ (for c quark). 
The combined values are $A_{FB}^{b} = -0.28 \pm 0.15$, $A_{FB}^{c} = -0.35
\pm 0.09$. The combined results are also shown in Figure \ref{fig:afb}.

\begin{table}[h]
  \caption{Differential cross sections for $e^{+}e^{-} \rightarrow c\bar{c}$ 
and $b\bar{b}$ processes.}
  \label{tab:cross-section}
  \begin{center}
     \begin{tabular}{ccc}
      \hline \hline
      $\cos\theta$ bin & $\frac{1}{\sigma_{had}}
			  \frac{d\sigma_{c\bar{c}}}{d\cos\theta}$ &
         $\frac{1}{\sigma_{had}}\frac{d\sigma_{b\bar{b}}}{d\cos\theta}$\\ 
      \hline
      -0.8 $\sim$ -0.6 & 0.153$\pm$0.113& 0.088$\pm$0.036\\
      -0.6 $\sim$ -0.4 & 0.238$\pm$0.049& 0.073$\pm$0.020\\
      -0.4 $\sim$ -0.2 & 0.101$\pm$0.038& 0.074$\pm$0.016\\
      -0.2 $\sim$  0.0 & 0.200$\pm$0.047& 0.046$\pm$0.014\\
       0.0 $\sim$  0.2 & 0.192$\pm$0.044& 0.047$\pm$0.015\\
       0.2 $\sim$  0.4 & 0.119$\pm$0.036& 0.038$\pm$0.016\\
       0.4 $\sim$  0.6 & 0.148$\pm$0.044& 0.062$\pm$0.018\\
       0.6 $\sim$  0.8 &-0.057$\pm$0.122& 0.037$\pm$0.025\\
      \hline \hline
    \end{tabular}\\
  \end{center}
\end{table} 

\section{Conclusion}
\label{sec:conclusion}

We have studied inclusive muon events and extracted 
the cross-sections and charge asymmetries of b and c-quark using 
all of the data collected with 
the TOPAZ detector at $\sqrt{s} = 58$ GeV. A total of 1,328 inclusive muon 
events were selected from 29,561 hadronic events with an  
integrated luminosity of 273 pb$^{-1}$. 
To improve the accuracy of the simulation for pion punch-throughs, 
we tuned the parameters in the Monte-Carlo simulation, using 
1,202 pions from $\tau^{\pm} \rightarrow \pi^{\pm}\pi^{\pm}\pi^{\mp}\nu$, 
and $\tau^{\pm} \rightarrow \rho^{\pm}\nu, \rho^{\pm} \rightarrow \pi^{\pm}
\pi^{0}$ in $e^{+}e^{-} \rightarrow \tau^{+}\tau^{-}$ reaction. 
The measured ratio, $R_{q\bar{q}}$, of the cross section 
for $q\bar{q}$ production to the total hadronic cross section 
and the forward-backward asymmetry, $A_{FB}^{q}$, of b and c 
quark are $R_{b\bar{b}} = 0.13\pm0.02(stat)\pm0.01(syst)$, $R_{c\bar{c}} 
= 0.36\pm0.05(stat)\pm0.05(syst)$, $A_{FB}^{b} = -0.20\pm0.16(stat)\pm
0.01(syst)$ and $A_{FB}^{c} = -0.17\pm0.14(stat)\pm0.02(syst)$, respectively. 
The standard-model prediction for those parameters 
with 12\% $B-\bar{B}$ mixing ($\chi_{B}$) are $R_{b\bar{b}} = 0.13$, 
$R_{c\bar{c}} = 0.30$, $A_{FB}^{b} = -0.43$,$A_{FB}^{c} = -0.48$, respectively.
The measured $R_{b\bar{b}}$ and $R_{c\bar{c}}$ agree with the standard-model
predictions, though the measured $A_{FB}^{b}$ and $A_{FB}^{c}$ are smaller 
by $2.2\sigma$ and $1.5\sigma$, respectively.

\noindent{\em Acknowledgments.}
\smallskip \\
\hspace*{12pt}We wish to express our deep thanks to the TRISTAN 
accelerator group for operating the TRISTAN accelerator excellently
for more than eight years. 
We are grateful to all of the engineers and technicians at KEK 
and the other collaborating institutions: H. Inoue, N. Kimura, K. Shiino, 
M. Tanaka, K. Tsukada, N. Ujiie, and H. Yamaoka.

\end{document}